\documentclass[5p,times,sort&compress]{elsarticle}
\PassOptionsToPackage{table,xcdraw,x11names,dvipsnames,table}{xcolor}
\usepackage{todonotes, xspace}
\usepackage{moreverb,url}
\usepackage{graphicx}
\usepackage{array}
\usepackage{multirow}
\usepackage{booktabs}
\usepackage{paralist}
\usepackage[most]{tcolorbox}
\usepackage{colortbl}
\usepackage{xcolor}
\usepackage{balance}

\AtBeginDocument{
  \definecolor{pdfurlcolor}{rgb}{0,0,0.6}
  \definecolor{pdfcitecolor}{rgb}{0,0.6,0}
  \definecolor{pdflinkcolor}{rgb}{0.6,0,0}
  \definecolor{light}{gray}{.85}
  \definecolor{vlight}{gray}{.95}
}

\usepackage[colorlinks,bookmarksopen,bookmarksnumbered,citecolor=red,urlcolor=red]{hyperref}
\usepackage{tabularx,ragged2e,rotating}

\newcommand\BibTeX{{\rmfamily B\kern-.05em \textsc{i\kern-.025em b}\kern-.08em
T\kern-.1667em\lower.7ex\hbox{E}\kern-.125emX}}

\newcommand{\eg}{e.g.,\xspace}
\newcommand{\ie}{i.e.,\xspace}
\newcommand{\pp}[1]{\vspace{6pt}\noindent\textbf{\emph{#1.}}\xspace}

\newcommand{\revised}[1]{{\color[rgb]{0,0,0}#1}}

\journal{arXiv}
\begin{document}

\begin{frontmatter}
\title{A Terminology for Scientific Workflow Systems}

\tnotetext[]{This manuscript has been authored in part by
    UT-Battelle, LLC, under contract DE-AC05-00OR22725 with the US
    Department of Energy (DOE). The US government retains and the
    publisher, by accepting the article for publication, acknowledges
    that the US government retains a nonexclusive, paid-up,
    irrevocable, worldwide license to publish or reproduce the
    published form of this manuscript, or allow others to do so, for
    US government purposes. DOE will provide public access to these
    results of federally sponsored research in accordance with the DOE
    Public Access Plan
    (\url{http://energy.gov/downloads/doepublic-access-plan}).}

\author[ORNL]{Frédéric Suter\corref{cor1}}           \ead{suterf@ornl.gov}
\author[UCSD]{Tainã Coleman\corref{cor1}}            \ead{t1coleman@ucsd.edu}
\author[UCSD]{{\.I}lkay Altinta\c{s}}   \ead{ialtintas@ucsd.edu}
\author[BSC]{Rosa M. Badia}             \ead{rosa.m.badia@bsc.es}
\author[AGH]{Bartosz Balis}             \ead{balis@agh.edu.pl}
\author[Uchicago]{Kyle Chard}           \ead{chard@uchicago.edu}
\author[UTorino]{Iacopo Colonnelli}     \ead{iacopo.colonnelli@unito.it}
\author[ISI]{Ewa~Deelman}               \ead{deelman@isi.edu}
\author[Seqera]{Paolo Di Tommaso}       \ead{paolo@seqera.io}
\author[Innsbruck] {Thomas Fahringer}   \ead{Thomas.Fahringer@uibk.ac.at}
\author[UMan]{Carole Goble}             \ead{Carole.Goble@manchester.ac.uk}
\author[PPPL]{Shantenu Jha}             \ead{shantenujha@acm.org}
\author[Illinois]{Daniel S. Katz}       \ead{d.katz@ieee.org}
\author[DUE]{Johannes Köster}           \ead{johannes.koester@uni-due.de}
\author[UBerlin]{Ulf~Leser}             \ead{leser@informatik.hu-berlin.de}
\author[ORNL]{Kshitij Mehta}            \ead{mehtakv@ornl.gov}
\author[NIWA]{Hilary Oliver}            \ead{hilary.oliver@niwa.co.nz}
\author[LLNL]{J.-Luc Peterson}          \ead{peterson76@llnl.gov}
\author[PSI]{Giovanni Pizzi}            \ead{giovanni.pizzi@psi.ch}
\author[LLNL]{Loïc Pottier}             \ead{pottier1@llnl.gov}
\author[BSC]{Raül Sirvent}              \ead{Raul.Sirvent@bsc.es}
\author[ORNL]{Eric Suchyta}             \ead{suchytaed@ornl.gov}
\author[NotreDame]{Douglas~Thain}       \ead{dthain@nd.edu}
\author[ORNL]{Sean R. Wilkinson}        \ead{wilkinsonsr@ornl.gov}
\author[ANL]{Justin M. Wozniak}            \ead{woz@anl.gov}
\author[ORNL]{Rafael Ferreira da Silva} \ead{silvarf@ornl.gov}

\address[ORNL]{Oak Ridge National Laboratory, TN, USA}
\address[UCSD]{University of California, San Diego, CA, USA}
\address[BSC]{Barcelona Supercomputing Center, Barcelona, Spain}
\address[AGH]{AGH University of Krakow, Krakow, Poland}
\address[Uchicago]{University of Chicago, Chicago, IL, USA}
\address[UTorino]{University of Torino, Torino, Italy}
\address[ISI]{Information Sciences Institute, University of Southern California, Marina del Rey, CA, USA}
\address[Seqera]{Seqera Labs, Barcelona, Spain} 
\address[Innsbruck]{University of Innsbruck, Institute of Computer Science, Innsbruck, Austria}
\address[UMan]{University of Manchester, Manchester, United Kingdom}
\address[PPPL]{Rutgers University-New Brunswick; Princeton Plasma Physics Laboratory; Princeton University, NJ, USA}
\address[Illinois]{NCSA \& School of Computing and Data Science \& iSchool, University of Illinois Urbana-Champaign, IL, USA}
\address[DUE]{University of Duisburg-Essen, Essen, Germany}
\address[UBerlin]{Institute for Computer Science, Humboldt-Universität zu Berlin, Berlin, Germany}
\address[NIWA]{National Institute of Water and Atmospheric Research, Wellington, New Zealand}
\address[LLNL]{Lawrence Livermore National Laboratory, Livermore, CA, USA}
\address[PSI]{PSI Center for Scientific Computing, Theory and Data, Villigen, Switzerland}
\address[NotreDame]{University of Notre Dame, Notre Dame, IN, USA} 
\address[ANL]{Argonne National Laboratory, Lemont, IL, USA}

\cortext[cor1]{Corresponding authors}

\begin{abstract}
The term ``scientific workflow'' has evolved over the last two decades to encompass a broad range of compositions of interdependent compute tasks and data movements. It has also become an umbrella term for processing in modern scientific applications. Today, many scientific applications can be considered as workflows made of multiple dependent steps, and hundreds of workflow systems have been developed to manage and run these scientific workflows. However, no turnkey solution has emerged from the field to address the diversity of scientific processes and the infrastructure on which they are supposed to be implemented. Instead, new research problems requiring the execution of scientific workflows with some novel feature often lead to the development of an entirely new workflow system. A direct consequence of this situation is that many existing workflow management systems (WMSs) share some salient features, offer similar functionalities, and can manage the same categories of workflows but at the same time also have some distinct capabilities that can be important for specific applications. This situation makes researchers who develop workflows face the complex question of selecting a WMS. This selection can be driven by technical considerations, to find the system that is the most appropriate for their application and for the computing and storage resources available to them, or other factors such as reputation, adoption, strong community support, or long-term sustainability. To address this problem, a group of WMS developers and practitioners joined their efforts to produce a community-based terminology of WMSs. This paper summarizes their findings and introduces this new terminology to characterize WMSs. This terminology is composed of fives axes: workflow structure and characteristics, composition, orchestration, data management, and metadata capture. Each axis comprises several concepts that capture the prominent features of WMSs. Based on this terminology, this paper also presents a classification of 23 existing WMSs according to the proposed axes and terms.

\end{abstract}


\end{frontmatter}

\section{Introduction}

The concept of \emph{workflows}, i.e., the execution of orchestrated and repeatable patterns of activity, dates back to the early 1900s when the engineering and manufacturing community introduced one of the earliest examples of procedural workflow: the Ford assembly line adopted by automobile manufacturers to this date. Workflows has been used to model, analyze, and improve business processes, using tools such as flow charts, functional flow block diagrams, or control flow diagrams~\cite{bpm}. The database community has also used workflows to address the challenges of managing large datasets~\cite{wiener}. The capacity to describe and orchestrate such complex applications popularized workflows across multiple scientific domains. The term \emph{scientific workflow} itself was introduced in 1996~\cite{wainer1996scientific, sheth} to differentiate this specific type of workflow from the business and automation pipelines that inspired them. As scientific workflow may designate processes that go beyond science to cover more broadly defined research activities, we opted for the use of the term \emph{workflow} in the remainder of this article with the following all-encompassing definition:

\smallskip
\pp{Definition} A workflow is a structured sequence of computational tasks or activities that achieve a research or analytical objective. Workflows define the flow of work, including the order of steps, the data and control dependencies between them, and the rules governing their execution. Modern workflows extend beyond traditional directed acyclic graphs to encompass dynamic, adaptive, and interactive processes that may include cycles, branches, and human-in-the-loop components. They span diverse domains, including scientific research, engineering, humanities, and business, and bridge heterogeneous computing environments from edge devices to high-performance computing (HPC) facilities and cloud infrastructure.
\smallskip

Over the past decades, workflows have become the predominant format for describing complex, multi-step, multi-domain scientific applications~\cite{badia2017workflows}. To manage the composition, planning, orchestration, and automation of the efficient execution of such workflows on powerful and often distributed compute infrastructures, a wide variety of workflow management systems (WMSs) have been proposed~\cite{wms-list}. However, domain researchers who develop workflows and want to rely on a WMS to execute them often face the complex question of selecting a particular WMS. This selection can be driven by technical considerations, such as finding the most appropriate system for their application and for the computing and storage resources available to them, or factors such as reputation, adoption, community support, or long-term sustainability. The main reasons for this challenge are that no single ideal turnkey solution has emerged from the field to address the diversity of scientific processes and the heterogeneity of possible execution environments (both in terms of hardware and software). Instead, new research problems or new computer technologies related to the execution of workflows often lead to the development of an entirely new workflow management system.   

\begin{figure*}[!ht]
    \centering
    \includegraphics[width=.97\linewidth]{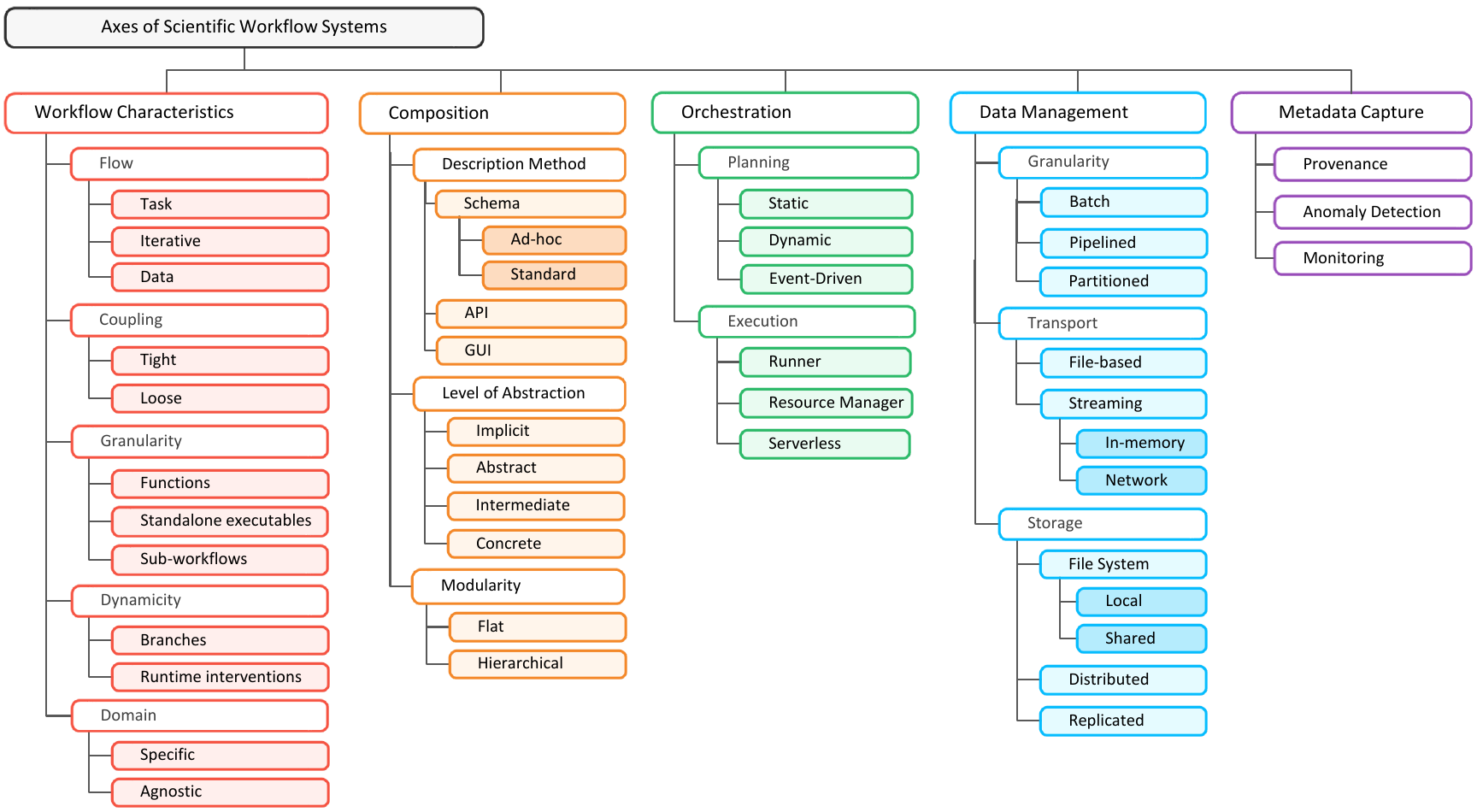}
    \caption{Five axes that categorize workflows and workflow management systems, with each axis further delineated into corresponding terms and sub-terms to provide a structured and detailed terminology.}
    \label{fig:axes}
\end{figure*}

A direct consequence of this situation is that many existing WMSs share some salient features, offer similar functionalities, and can manage the same categories of workflows, but often also have distinct features tailored for specific types of problems. This has been highlighted by different efforts to create taxonomies and characterizations of workflows and WMSs~\cite{Yu_taxo05, silva-fgcs17, deelman-fgcs09, liu-grid15, bahsi-2007, kiran2023criteria, Ahmad2021, POOLA2017285, cruz2009}. These efforts can help to provide workflow developers with some guidance when trying to select the appropriate tool to develop and execute their workflows, but they are also notoriously incomplete and quickly outdated in a fast-moving field. Consequently, decisions for specific systems are very often based on social aspects as much or more than on technical ones (e.g., previous experience in the community, word-of-mouth, comments in web forums, personal evaluation of a few known systems). Inspired by the work of the in situ processing community for data visualization and analysis systems~\cite{in-situ-term}, we propose in this article to go beyond a traditional taxonomy and develop a consistent terminology to describe WMSs. \revised{While it shares similarities with and builds on existing taxonomies, the driving principle of this effort was to determine terms that consensually describe the high-level features of workflows and WMSs, rather than categorizing systems based on implementation details}.

To this end, we gathered a group of workflow system developers and workflow practitioners, all members of the Workflows Community Initiative~(WCI)~\cite{WCI}, and followed a process similar to that in~\cite{in-situ-term} to create a strong terminology for WMSs. This paper synthesizes the discussions initiated during the different editions of the Workflows Community Summit~\cite{wci-roadmap,da2021workflows, da2021workflows1, da2022workflows, wcs2022, wcs2024}, which led to the writing of this paper. The main contribution of this paper is the identification of five axes to characterize WMSs (Figure~\ref{fig:axes}). Each axis comprises a series of concepts that capture the most salient features of WMSs. Based on the proposed terminology, our group analyzed 23 actively developed WMSs that are part of the WCI to determine which combination of terms can define each of them.

The remainder of this paper is organized as follows. Section~\ref{sec:axes} defines the proposed five axes to describe a workflow system. Section~\ref{sec:survey} reviews the 23 selected WMSs and classifies them according to the proposed axes and terms. Section~\ref{sec:process} describes the process followed by members of the WCI to produce this terminology \revised{and Section~\ref{sec:related-work} discusses previous efforts to establish taxonomies of WMSs}. Finally, Section~\ref{sec:ccl} summarizes our work.

\section{Axes of Scientific Workflow Systems}
\label{sec:axes}

WMSs often consist of subsystems that handle specific aspects of workflow management, such as resource allocation, task scheduling, or, data management. A WMS coordinates these subsystems to ensure efficient and robust execution. Additionally, characterizing a given workflow system requires considering the characteristics of the workflows it can support, as it often influences the design of the system.
The primary goal of defining this terminology is to help scientists navigate the wide range of available tools~\citep{wms-list} and better express their computational needs. To achieve this, we identified five key axes to describe a workflow system:

\begin{compactitem}
    \item {\bf Workflow Characteristics:} This axis examines fundamental organizational aspects that impact how workflows operate and adapt. Specifically, it examines how execution is driven (by tasks or data), the level of complexity of individual components, the nature of dependencies between these components, and the ability to modify execution paths at runtime. These structural elements significantly influence how WMSs optimize resource use and performance.
    
    \item {\bf Composition:} This axis addresses how workflows are defined, organized, and configured by WMSs. It explores the methods used to describe workflows, the level of detail required in these descriptions, and how complex workflows can be shaped from simpler components. This axis helps to understand how accessible and flexible different WMSs are for users with varying technical backgrounds.
    
    \item {\bf Orchestration:} This axis covers the implementation and execution management approaches for workflow components. It analyzes different methods for launching and coordinating tasks, from direct execution to more sophisticated approaches that leverage distributed resources, event-based triggers, or cloud services. These orchestration strategies determine how efficiently workflows use available computing infrastructure.
    
    \item {\bf Data management:} This axis focuses on how data is handled throughout the workflow lifecycle. It characterizes methods for moving data between workflow components, approaches to storing data at different stages, and techniques for optimizing data access patterns. These data management strategies significantly affect workflow performance, especially for data-intensive applications.
    
    \item {\bf Metadata capture:} This axis explores additional contextual information collected during workflow execution. It covers methods for tracking workflow execution state, documenting provenance, monitoring performance, and detecting anomalies. These capabilities ensure that workflows can be reliably executed, optimized,  debugged, and reproduced.
\end{compactitem}

Figure~\ref{fig:axes} provides an overview of the terms used for each axis, and this section describes these terms in more detail. With this terminology, we can describe a WMS based on a selection of specific terms for each of the five identified axes. As sub-terms within an axis are not mutually exclusive, a WMS may be classified by a combination of sub-terms.  

\subsection{Workflow Characteristics}
\label{sec:structure-and-characteristics}

The first axis is more focused on the type of workflows a workflow system can manage than on the characteristics of a system itself, in other words, on {\it what} the workflow does. The large number of existing WMSs~\cite{wms-list} indicates that there is no ``one-size-fits-all" solution despite standardization and interoperability efforts~\cite{cwl}. In fact, the design and implementation of workflows is significantly influenced by structural aspects that are crucial to their efficiency, scalability, and adaptability. In this section, we characterize broad classes of workflows according to these defining features. 

A prominent feature of a workflow is its {\bf flow}, which has a direct impact on how WMSs optimize workflow execution. When workflow components receive inputs, process them, generate outputs, and then terminate, the workflow structure is defined by the composition of these \textbf{tasks}. WMSs are then responsible for orchestrating their execution, respecting their flow and control dependencies. They will also implement optimization strategies to improve workflow performance, such as minimizing the makespan or communication of the workflow. The different tasks that make up a workflow can also be executed multiple times in an {\bf iterative} way. At each iteration, tasks are executed, terminated, and then wait to be invoked again. The structure and execution of the workflow can also be driven by the {\bf data} flowing through the workflow components. These components are data operators that remain alive while there is data to process. In that case, WMSs aim to maximize the throughput of the workflow. 

The structure of workflows also differs by the {\bf granularity} of their individual tasks. Some workflows can compose some {\bf function} calls to perform complex processing tasks. To some extent, a script or a program can be seen as a workflow and a runtime system as a workflow system. The most common definition of a workflow is a composition of {\bf standalone executables}, which aggregate multiple functions calls to perform complex computations on a set of inputs and produce a set of outputs. With the increase in scale and complexity of computational problems, it is now common to express workflows as a hierarchical and modular composition of {\bf sub-workflows}.

Another defining feature of workflows is the {\bf coupling} of the tasks that compose them. This term defines the dependencies and interactions between the different tasks. The {\bf tight} coupling of some tasks indicates that these tasks must be executed concurrently, being co-located on the same computing resources or running on different sets of processors. This is often caused by periodic data exchanges between tasks while they run. Conversely, a {\bf loose} coupling of tasks does not impose any constraint on the concurrent execution of tasks, giving more flexibility to the WMS when scheduling the workflow. 

The {\bf dynamicity} of a workflow indicates its ability to modify its structure during its execution. Dynamic workflows can comprise several {\bf conditional branches} that are activated or not depending on the realization of a predefined condition or triggered by an external event. Such conditions can be related to changes observed in the processed datasets (\eg a variable reaching a certain threshold, the convergence of an iterative process is reached),  to changes in the status or availability of compute, network, or storage resources, or to time-related events (\eg it is too late to process a given execution path). Such conditional branches allow workflows and WMSs to efficiently react to changes and foster more robust, efficient, and flexible executions. A second type of dynamic behavior found in workflows is when a {\bf runtime intervention} is needed. In that case, the workflow system gives the control back to the user who started the workflow or to an automated external decision process. Such interventions at runtime can modify the initial execution plan of a workflow in different ways (\eg rerun certain tasks or an entire sub-workflow, modify task configuration, cut a given path or start exploring a new path, or trigger the early termination of the entire workflow).

Finally, it is possible to distinguish WMSs with respect to the {\bf domain} they serve. Some systems are deeply rooted in a scientific community and thus mainly target domain-\textbf{specific} workflows, while others are more application-\textbf{agnostic}.

\subsection{Composition}
\label{ssec:composition}

Composition refers mainly to how a workflow system allows its users to describe the different components of the workflow, their configuration and input parameters, and the data and control dependencies between these components. This axis also covers the coupling between the description of the workflow itself and that of the targeted hardware and software infrastructure on which to execute the workflow. 

We identified three subcategories of {\bf description methods} to compose a workflow. The first, \textbf{schema}, refers to the case where the workflow is described in a text file, using a specific format (\eg XML, JSON, YAML, or a domain-specific language) and syntax. We further decompose this category to distinguish that the syntax used by a WMS is \textbf{ad-hoc}, meaning that it can only be understood by this particular WMS, or part of a common \textbf{standard} shared by multiple WMSs, such as the Common Workflow Language~(CWL)~\cite{cwl}, the Interoperable Workow Intermediate Representation (IWIR)~\cite{iwir}, or WfFormat~\cite{wfcommons}. Note that supporting a description standard may not always be possible, for instance, when a WMS implements significant features that cannot be easily expressed in the standard. The second subcategory includes WMSs that expose an {\bf API} to describe workflows. This API builds on or extends one or more popular programming languages (\eg Python, C++) or a text templating engine (\eg jinja) to leverage loops and conditional statements and allow users to describe their workflows in a more compact and flexible way. The third subcategory corresponds to WMSs that rely on a \textbf{graphical user interface~(GUI)}. 

Workflow composition can also be defined by the \textbf{level of abstraction} of the description provided by the user. A \textbf{high-level abstract} composition will only focus on describing the logical structure of the task graph, a generic description of the data flowing through the workflow, and the amount of resources required by each component. This abstract description is generally independent of a specific instance of the workflow (\ie that specifies all input parameters and component configuration parameters) and of a specific target computing and storage infrastructure. The advantages of an abstract workflow composition are that it favors the reusability and portability of the workflow. However, it requires more effort from users or the workflow system to execute a specific instance on a specific infrastructure. 

Some systems have an \textbf{intermediate-level abstract} composition. They allow for a high-level workflow description while requiring some execution details from the users. Systems with intermediate-level abstraction provide users with a balance between automation and manual fine-tuning, which can be advantageous when the application requires a higher level of execution control. This comes at the cost of lower portability when compared to fully abstract systems and possible performance trade-offs (e.g.,  the responsibility of allocation optimization falls on the users in these systems). 

Conversely, a \textbf{concrete} composition is more closely related to an instance and an infrastructure. All parameters are specified in the description, and the workflow can be deployed and run directly from it. Note that some WMSs allow to factor infrastructure related information as a separate description, allowing users to port a workflow from one infrastructure to another without changing the high- or intermediate-level abstract composition of the workflow itself. 

When an API is used to describe a workflow, the composition is {\bf implicit} as the workflow's structure is derived from the composition of the different function calls made by the user, or from metadata attached to a dataset to process, indicating for instance which files are needed and in what way.   

Finally, we also distinguish composition methods with regard to their \textbf{modularity}. With the evolution of scientific applications from relatively simple workflows (\eg data processing pipelines or fan-out/fan-in execution patterns for ensemble runs) to more complex workflows composed of interconnected sub-workflows (\ie workflow of workflows), the composition methods exposed by WMSs are also evolving from a \textbf{flat} description of a set of components to a more \textbf{hierarchical} description that enables modular and scalable design. This shift allows for better management of large-scale applications, where individual sub-workflows can be developed, tested, and optimized independently before integration. It also allows researchers to create new complex data analysis workflows by composing existing workflows developed in their community. However, such hierarchical composition introduces new challenges, such as dealing with an increased orchestration complexity, handling dependencies across nested workflows, and efficiently managing resource allocation. To address these, WMSs provide features such as parameterized workflow components and reusable templates that facilitate modular workflow design while maintaining scalability.

\subsection{Orchestration}
\label{ssec:orchestration}

Orchestration refers to the method(s) employed by a workflow system to deploy, schedule, and execute the computational components of a workflow. In this section, we focus on the general functional features of WMSs rather than the specific technical details of their implementations. For instance, we leave optimization techniques, such as advanced, performance-oriented scheduling and resource allocation techniques and algorithms, out of the scope of this axis. However, we still consider it important to classify WMSs into three broad categories related to execution {\bf planning}. Some systems impose a {\bf static} planning of the workflow execution, i.e., all the decisions about when and where each task composing the workflow is executed must be taken before the execution starts. Conversely, some systems can make or adapt scheduling and resource allocation decisions during the execution of the workflow, hence implementing a {\bf dynamic} planning strategy. (Note that certain systems implementing static planning may emulate dynamic planning through hierarchical workflows.) The third category encompasses WMSs that do not plan the workflow execution in advance but rather let the execution react to specific events and/or conditions that occur at runtime. In such {\bf event-driven} execution, when a trigger condition or event is met, the workflow system automatically initiates subsequent, usually predefined, actions such as starting new tasks, notifying users, and adjusting the resource allocation. This type of automation minimizes manual intervention, making the orchestration less error-prone.

We identified three categories for the actual {\bf execution} of the tasks that compose a workflow. WMSs might use one or more orchestration methods (see Table~\ref{tab:workflow_systems}) to execute a workflow. The \textbf{runner} orchestration method refers to WMSs that are fully responsible for the acquisition of computing and storage resources and the management of the individual tasks that compose a workflow. It connects the high-level workflow definition (i.e., its composition, see Section~\ref{ssec:composition}) to the available resources. A runner system ensures that tasks execute in the correct order, respecting their pre-defined control and flow dependencies. It oversees the life cycle of a task from the time it is dispatched and monitors it until it is completed according to its specifications. 

Other WMSs delegate resource allocation and part of the management of the execution of individual tasks to a {\bf resource manager}. This orchestration method is typically used in HPC systems where the allocation of compute nodes is handled by a batch scheduler, or cloud systems, where container orchestration systems are used. The interactions between the workflow system and the underlying resource managers encompass ordering queue of jobs to execute in an ensemble, controlling the release of limited quantities of tasks or data to not overwhelm the underlying execution system, or implementing a \emph{pilot job}~\cite{10.1145/3177851} mechanism to reduce the queuing overhead caused by scheduling and executing tasks independently by grouping them within the pilot allocation.

The last orchestration method relies on a \textbf{serverless} execution of tasks. This refers to a cloud-based model in which the responsibility for infrastructure management, allocation scaling, and job execution is entirely delegated to a cloud service provider. A key distinction of this model is that the user or WMS must first define one or more functions along with all of their software dependencies, and then the WMS may execute those functions to carry out the workflow.  The cloud platform takes care of the provisioning and server management, abstracting the underlying computing and storage infrastructure entirely. In some cases, it can be the most cost effective orchestration method as users are usually only charged based on the actual usage of computing resources rather than maintaining servers always on, even when idle.

\subsection{Data Management}

The data management axis characterizes the way WMSs transport, store, and manage the lifecycle of one of the key components of scientific workflows: data. Before detailing the different categories and terms related to data management, we make an important distinction between two types of data, as the way a workflow system manages each of them may differ. \textbf{Input/output} data respectively refer to the data needed at the beginning of the workflow and to the final outcomes of its execution, while \textbf{intermediate} data denotes every piece of data that did not exist before the beginning of the workflow and will not be kept after the end of this execution.  

A first way to distinguish WMSs according to how they manage data is to consider the {\bf granularity} at which these systems handle data management operations. A common approach followed by many WMSs is to consider the data operations of a workflow component at the granularity of a {\bf batch}: all the needed input data are consumed before performing computations and all the output data is produced, and made available to subsequent components in the workflow, at the end of these computations. 

Another approach is to consider a {\bf pipelined} granularity in which workflow components periodically produce and/or consume individual records during their entire lifecycle. This is typically used to manage in situ processing workflows~\cite{10.1145/3592979.3593420}, where analysis and visualization components are loosely coupled to a main data producer (i.e., a numerical simulation). In such workflows, data is consumed as it is produced, in opposition to a {\it post-hoc} approach in which analyses or visualization happens once the full dataset has been generated. 

A third intermediate granularity is to consider data as {\bf partitioned}, i.e., divided in groups of individual records, and to transfer these partitions across the workflow. This approach is particularly useful when individual records are small. Considering them individually would be very latency-sensitive and could negatively impact performance. 

A second way to differentiate WMSs is by how they {\bf transport} data from one workflow component to another. Again, a common approach is to rely on {\bf file-based} transport, in which a workflow component that produces intermediate data will write them into a file(s) on a storage system. In contrast, a workflow component that consumes intermediate data will read it from file(s). An alternate approach is to directly {\bf stream} intermediate data between components. Depending on the respective allocations of the producing and consuming components, it is possible to further refine these two broad approaches. 

For WMSs that rely on the file-based transport approach, we can further distinguish them according to the {\bf storage} they use. When workflow components are co-located on the same compute node, the workflow system can leverage the existence of a {\bf local file system}, while when components are allocated to different nodes of the same compute cluster or to different clusters of the same computing facility, it will have to rely on a {\bf shared file system}.  Commonly used in collaborative or high-performance computing environments, shared file systems correspond to  a centralized model where data is accessible by multiple systems or nodes simultaneously. They bring several advantages when executing workflows, such as simple and collaborative access to a unified storage space or good cost efficiency. They also come with different challenges, such as data consistency, performance bottlenecks, scalability, or security, that a WMS will have to face, and may address.  In the extreme case where the execution of a workflow is distributed over multiple computing facilities, this approach can leverage a {\bf distributed storage space}. This involves managing and storing data across multiple local and/or remote systems, enabling scalability, load balancing, resilience, and flexibility. Although it can resolve some issues of shared file systems, data consistency and security challenges persist. Furthermore, the management of such systems can be very complex, and data accesses may suffer from high latencies. An alternative approach in that case would be that the data-producing workflow components running in a given facility create one or several additional transfer tasks to send data to each their its data-consuming successors that run in another facility. Another common practice in distributed and shared systems targeted by WMSs is the use of \textbf{replicated storage}, which focuses on creating redundant copies of data to improve reliability, availability, and resilience. Such as the aforementioned storage solutions, replicated storage struggles with data consistency and complex data management, not to mention the increased storage costs and the write overhead created every time data needs to be updated. 

For the stream-based transport approach, when the producer and consumer are co-located on the same node, data transport can be carried out {\bf in-memory} through a shared address space. Otherwise, it implies a {\bf network} communication between the nodes that respectively hosts the data producer and consumer.

\subsection{Metadata Capture}

The last axis of the terminology refers to the different categories of contextual information, or metadata, captured by WMSs during a workflow execution. Metadata constitute a critical layer of information that describes, tracks, and contextualizes workflow aspects such as inputs/outputs, parameters, and dependencies. Through the extra information, the workflow engine can decide on the execution order based on the dependency information, parallelize tasks, and schedule resources according to the needs of the task. Therefore, metadata enables efficient orchestration, automation, long-term data management, and resource optimization. By capturing descriptive execution logs and storing full context results, metadata can improve troubleshooting, debugging, and responsiveness. Overall, it can ensure scientific integrity, reproducibility, and reliability throughout the workflow's lifecycle. 

Workflows are typically large and complex applications designed for execution in distributed systems. Given their role in critical research and high-impact projects, the ability to reproduce results enables others to validate the findings, build upon previous work,  and promote collaboration to further scientific discovery.

A specific type of metadata is \textbf{provenance} data which can be further decomposed into prospective and retrospective provenance data. Prospective provenance corresponds to maintaining detailed information about the workflow design and structure, the configuration of the workflow system and the underlying computing and storage infrastructure, and the specific algorithms to be used and their parametrization. Prospective provenance is essential to facilitate reproducibility, especially for complex applications such as workflows~\cite{leo2024recording}. Retrospective provenance data corresponds to what actually happened to the data processed by a workflow and captures everything related to a specific execution. It is usually extracted from execution logs to keep track of the data lineage (i.e., generation, transformation, and usage) and timestamps and runtime details. Retrospective provenance is particularly useful for detecting any deviation from the expected execution plan and is often used for debugging purposes.

Another type of metadata captured during workflow executions is \textbf{monitoring} of data, which comes from processes that oversee the workflow execution in real time. The data generated by monitoring provides critical insight into performance, resource utilization, and potential bottlenecks. WMSs can leverage it to dynamically reconsider an initial execution plan by modifying resource allocations or scheduling decisions. The monitoring data can also be analyzed by researchers after a workflow execution to optimize the description of the workflow itself to improve its efficiency.

The final category on this axis is related to \textbf{anomaly detection}~\cite{RAGHAVAN2025107608}. We consider that a workflow management system supports anomaly detection if it captures metadata that can be used to implement fault tolerance mechanisms. These mechanisms vary in sophistication: Some systems terminate execution and display an error message, while others complete the execution but log warnings about potentially incorrect data resulting from unexpected behavior. There are even systems that can distinguish between anomalies that can be handled automatically (e.g., task retries or by an optional branch from a task-failed trigger) and anomalies that the workflow is not designed to handle and thus require user intervention. In the latter case, the scheduler remains alive on a timeout in a ``stalled'' state, awaiting operator intervention.

\section{Surveying Existing Workflow Systems}
\label{sec:survey}

\revised{
This section considers {\bf 23 WMSs} that are part of the Workflows Community Initiative (WCI). This selection is motivated by the fact that the WCI focuses on \emph{actively developed} WMSs with a \emph{large user base}. We also ensured that the selection made was not limited to a specific research community, a narrow set of origin countries, or a certain category of supported workflows to avoid biases in the definition of our terminology.} Although this list represents only a small fraction of the vast number of existing WMSs~\cite{wms-list} and is thus far from being exhaustive, we believe that it is still representative of the diversity of the available systems. \revised{Moreover, this initial list of analyzed systems is not definitive nor intended to be limited to WMSs affiliated to the WCI. We plan to make this terminology available on the WCI website and broadly advertise its existence so that the list of WMSs mapped to the terminology continues to grow.}

\revised {For each WMS, we} analyze their published work and incorporate feedback from community efforts over the past four years. Table~\ref{tab:merged_workflow_classification} summarizes the type of workflows each system is able to execute, while Table~\ref{tab:workflow_systems} highlights the primary characteristics of each system according to the axes and terms summarized in Figure~\ref{fig:axes} and detailed in Section~\ref{sec:axes}. Table~\ref{tab:workflow_systems} also includes a column named \texttt{extensions}, which lists additional functionalities that WMSs can support beyond their default configurations. These extensions may include optional plugins, third-party integrations, or interoperability with cloud-based storage and computing resources.

\newcommand{\tabhead}[1]{\hfill \textbf{#1} \hfill\hfill}
\renewcommand{\arraystretch}{1.11}

\begin{table*}[ht!]
    \centering
    \small
    \rowcolors{2}{gray!25}{white}

    \begin{tabular}{l c | p{1.1cm} p{2.2cm} p{1.4cm} p{2.8cm} p{1.8cm}} 
        \toprule
        \multicolumn{2}{c|}{\bf Name} & \tabhead{Flow} & \tabhead{Granularity} & \tabhead{Coupling} & \tabhead{Dynamicity} & \tabhead{Domain} \\
        \midrule
        AiiDA        & \citep{aiida2020} & Task \newline Iterative & Sub-workflows \newline Executables \newline Functions & Loose & Branches \newline Runtime intervention & Agnostic\\
        AirFlow      & \citep{airflow} & Task & Executables & Loose & Branches & Agnostic\\
        Apollo       & \citep{apollo} & Task \newline{Data} \newline{Iterative} & Functions \newline{Sub-workflows} & Loose & Branches & Agnostic\\
        COMPSs     & \citep{pycompss} & Task \newline{Iterative} & Functions  \newline Sub-workflows  \newline Executables  & Loose & Branches  & Agnostic\\
        Cylc         & \citep{cylc} & Task \newline{Iterative} & Executables\newline Sub-workflows & Loose & Branches \newline Runtime intervention & Agnostic\\
        Dask         & \citep{dask} & Data & Executables & Tight & - & Agnostic\\
        EFFIS        & \citep{EFFIS2} & Data & Executables & Tight \newline Loose & - & Specific\\
        FireWorks    & \citep{fireworks} & Task & Sub-workflows & Tight & Branches & Agnostic\\
        Galaxy       & \citep{galaxy} & Data & Executables \newline Sub-workflows & Loose & Branches\newline Runtime intervention & Agnostic\\
        Globus Compute & \citep{funcx} & Data & Functions \newline{Executables} & Loose & - & Agnostic\\
        HyperFlow    & \citep{hyperflow} & Data & Functions \newline{Executables} & Loose & - & Agnostic\\
        Makeflow     & \citep{makeflow} & Data & Sub-workflows & Loose & - & Agnostic\\
        Merlin       & \citep{merlin} & Task \newline{Iterative} & Sub-workflows & Loose & - & Agnostic\\
        MLFlow       & \citep{mlflow} & Task \newline{Iterative} & Executables & Loose & - & Specific\\
        Nextflow& \citep{nextflow} & Data & Sub-workflows & Loose & Branches & Agnostic\\
        Parsl        & \citep{parsl} & Data & Sub-workflows & Loose & Branches & Agnostic\\
        Pegasus      & \citep{pegasus} & Data & Sub-workflows \newline Executables  & Loose & Branches & Agnostic\\
        Radical  & \citep{radical} & Task \newline{Iterative} & Functions & Tight & - & Agnostic\\
        Snakemake    & \citep{snakemake} & Task \newline{Iterative} & Sub-workflows \newline Executables \newline Functions & Loose \newline Tight & Branches & Agnostic\\
        StreamFlow   & \citep{streamflow} & Task \newline{Data} \newline{Iterative} & Sub-workflows \newline{Executables} & Loose & Branches  & Agnostic\\
        Swift/T      & \citep{swiftT} & Task \newline{Data} & Functions & Tight & Branches \newline Recursion & Agnostic\\
        TaskVine     & \citep{taskvine} & Task \newline{Iterative} & Functions  \newline{Executables} & Loose & - & Agnostic\\
        Toil         & \citep{toil} & Data & Sub-workflows & Loose & Branches & Agnostic\\
        \bottomrule
    \end{tabular}
    \caption{Classification of workflow management systems based on structure and characteristics.This classification represents the state at the time of publication, to the best of the authors knowledge. As many of the presented WMSs constantly evolve, we suggest the reader to explore their respective documentation to get an up-to-date view of their capabilities and characteristics.}
    \label{tab:merged_workflow_classification}
\end{table*}

\begin{table*}[ht!]
    \centering
    \scriptsize
    \rowcolors{2}{gray!25}{white}

    \begin{tabular}{p{1.15cm} | p{1.57cm} p{1.36cm} p{1.26cm} | p{1cm} p{1.12cm} | p{1.15cm} p{1.2cm} | p{1.12cm} | p{2.65cm}} 
        \toprule
         & \multicolumn{3}{c|}{\tabhead{Composition}} & \multicolumn{2}{c|}{\tabhead{Orchestration}} & \multicolumn{2}{c|}{\tabhead{Data Management}} & \tabhead{Metadata}& \\
        \tabhead{Name} & \tabhead{Description} & \tabhead{Abstraction} & \tabhead{Modularity} & \tabhead{Planning} & \tabhead{Execution} & \tabhead{Transport} & \tabhead{Storage} & \tabhead{Capture} & \tabhead{Extensions}  \\
        \midrule
        AiiDA        & API & Intermediate & Hierarchical & Dynamic & Runner & File-based & Shared & Anomaly \newline Provenance & Plugins \newline Caching \newline Fault tolerance \newline HPC execution\\
        AirFlow      & API & Intermediate & Flat & Static & Runner & Stream & Shared & Monitoring & Dynamic pipelines\\
        Apollo       & Ad-hoc Schema & Abstract & Hierarchical & Dynamic & Resource Manager  \newline Serverless & Stream & Distributed  & Monitoring & Container/serverless \newline{Multi-cloud} \newline{Edge/cloud} \newline{Multi-objective scheduling}\\
        COMPSs     & API & Intermediate & Flat \newline Hierarchical & Dynamic &  Resource Manager \newline Serverless & Stream \newline File-based & Local \newline Shared \newline Distributed  & Anomaly \newline Monitoring \newline Provenance & Adaptive resource \newline allocation \newline HPC scalable \newline Replicated storage \\
        Cylc         & Ad-hoc Schema \newline API/templating & Concrete & Flat \newline Hierarchical & Static \newline Event-driven & Runner \newline Resource-manager & File-based & Shared  & Anomaly \newline Provenance \newline Monitoring  & HPC Execution \newline Plugins \newline Config templating \\
        Dask         & API & Concrete & Flat & Dynamic & Runner & Stream & Shared \newline Distributed  & Anomaly \newline Monitoring \newline Metadata & Python Libraries \newline Cluster Management \newline GPU Accel.\\
        EFFIS        & API & Intermediate & Flat &  Dynamic & Resource Manager & Stream \newline File-based & Shared \newline Distributed \newline Replicated & Anomaly \newline Monitoring & \\
        FireWorks    & API \newline Ad-hoc Schema &  Intermediate & Hierarchical & Dynamic & Resource Manager & File-based & Shared \newline Replicated & Anomaly \newline Monitoring \newline Provenance & Multi-platform execution\\
        Galaxy       & GUI \newline Ad-hoc Schema & Concrete &  Flat & Event-Driven & Runner & Stream & Shared & Anomaly \newline Monitoring \newline Provenance & External Tools \newline Execution API\\
        Globus\newline Compute & API & Abstract & Hierarchical & Dynamic & Resource Manager \newline Serverless  & Stream \newline File-based & Shared & Anomaly \newline Monitoring & Distributed storage\\
        HyperFlow    & Ad-hoc Schema &  Intermediate &  Flat & Static \newline{Dynamic} & Runner & Stream & Shared \newline Distributed & Provenance & Replicated storage \newline Cloud Integration \newline Scalability\\
        Makeflow     & Standard \newline (Make) & Abstract &  Hierarchical & Static & Runner & File-based  & Shared \newline Replicated & Anomaly \newline Monitoring \newline Provenance & Distributed storage\newline HPC execution\\
        Merlin       & Ad-hoc Schema & Intermediate &  Hierarchical & Static & Runner & File-based & Shared \newline Distributed \newline Replicated  & Anomaly \newline Monitoring \newline Provenance & Cloud-native Support\\
        MLFlow       & API & Intermediate & Flat & Static & Runner  & File-based \newline Stream & Shared \newline Distributed & Monitoring & \\
        NextFlow     & Ad-hoc Schema & Abstract & Hierarchical & Dynamic & Runner & Stream \newline File-based & Shared \newline Distributed & Anomaly \newline Monitoring \newline Provenance & Replicated storage  \newline Container/Cloud Support \newline HPC execution\\
        Parsl        & API & Abstract & Hierarchical & Dynamic & Runner & Stream \newline File-based & Shared \newline Distributed & Anomaly \newline Monitoring & Replicated storage  \newline Dynamic Parallelization \newline Cloud/Grid Support\\
        Pegasus      & Ad-hoc Schema \newline API & Abstract &  Hierarchical & Static & Runner &  File-based & Shared \newline Distributed & Anomaly \newline Monitoring \newline Provenance & Replicated storage \newline Multi-level Scheduling\\
        Radical\newline  & API & Abstract & Hierarchical & Static & Resource Manager & File-based & Shared \newline Distributed  & Anomaly \newline Monitoring \newline Provenance  \newline Metadata & Replicated storage \newline Scalable\\
        Snakemake    & Ad-hoc Schema & Abstract & Flat \newline Hierarchical & Static \newline Dynamic \newline Event-driven & Runner & File-based & Shared \newline Distributed & Anomaly \newline Monitoring \newline Provenance & Plugins \newline Scripting integration \newline Software deployment integration \newline Interactive reporting\\
        StreamFlow   & Standard (CWL) & Abstract &  Hierarchical & Dynamic & Runner \newline{Resource Manager} & File-based & Distributed & Anomaly \newline Provenance & Replicated storage \newline Cloud Integration\\
        Swift/T      & Ad-hoc Schema & High-level & Flat & Dynamic  & Resource Manager & Stream \newline File-based & Shared & Anomaly \newline Monitoring & Local Storage~\cite{Hercules_2017} \newline AI/ML Control~\cite{EMEWS_2018} \newline Parallel Tasks~\cite{MPI_Launch_2019} \\
        Taskvine         & API & Intermediate & Flat & Dynamic & Resource Manager & File-Based & Shared \newline Distributed \newline Replicated  & Anomaly \newline Monitoring \newline Provenance \newline & Serverless \newline Autoscaling \newline HPC Execution \newline Recoverable storage\\
        Toil         & Standard\newline (CWL/WDL) & Abstract & Hierarchical & Static & Runner & Stream \newline File-based & Shared \newline Distributed  & Anomaly \newline Monitoring \newline Provenance & Replicated storage \newline Multi-Cloud Support\\
        \bottomrule
    \end{tabular}
    \caption{Categorization of various workflow systems with respect to their composition, orchestration, data management, and information capture. In addtion, the last column highlights exemplary extensions provided beyond this common terminology. This classification represents the state at the time of publication, to the best of the authors knowledge. As many of the presented WMSs constantly evolve, we suggest the reader to explore their respective documentation to get an up-to-date view of their capabilities and characteristics.}
    \label{tab:workflow_systems}
\end{table*}

\pp{Evolution of Workflow Characteristics}
The evolution of WMSs in the past two decades reflects significant changes in computational approaches. Initially predominantly task-driven, workflows have expanded to embrace data-driven processing pipelines with the rise of big data. Modern workflows now integrate both paradigms, particularly as AI becomes embedded in research, enabling complex analytical pipelines that respond dynamically to data while preserving the structured execution needed for reproducibility. The growing complexity of applications called for greater composability and modern WMSs now support hierarchical sub-workflows and iterative processes, which allows researchers to independently develop and optimize components before integration. These systems have also evolved to support more dynamic execution through conditional branches, runtime interventions, and adaptive processing. However, while technical capabilities continue to expand, the scientific domains supported by WMSs are often determined more by social dynamics than by technical limitations.

\pp{The Social Dynamics of Workflow System Selection}
While the technical characteristics described in our terminology provide a foundation for evaluating WMSs, the actual selection process in practice is often significantly based on social factors. Our community observations reveal that researchers frequently choose WMSs based not solely on technical merits, but on established social patterns and connections. When confronted with multiple technically viable options, scientists typically gravitate toward systems already in use by their immediate collaborators, departmental colleagues, or disciplinary communities. This preference for socially validated tools creates adoption groups within research domains and institutions. The perceived credibility of a workflow system is substantially enhanced when it appears in trusted publications or receives endorsements from respected colleagues. In addition, institutional knowledge transfer plays a crucial role, as existing expertise and support infrastructures significantly lower the barrier to adoption. These social dynamics create self-reinforcing adoption patterns that can sometimes override purely technical considerations, highlighting that workflow system selection exists within a complex socio-technical ecosystem where community practices, established knowledge bases, and trusted relationships often determine final choices. Nevertheless, this understanding emphasizes why developing a common terminology is particularly valuable, i.e., it provides a framework for discussing technical aspects objectively while acknowledging the legitimate influence of social factors on technology adoption.

\pp{Emerging Patterns in Modern Workflow Systems}
Several significant trends are reshaping the landscape of WMSs. The traditional schema-based approach to workflow composition is giving way to API-driven interfaces, reflecting broader programming paradigm shifts and resulting in less abstract, more programmatic workflow descriptions. This transition enables finer control over workflow execution while sometimes sacrificing portability across environments. Simultaneously, WMSs are increasingly addressing the need for dynamic execution capabilities, responding to growing demands from scientific applications that require adaptive runtime behaviors and conditional processing paths. Data management approaches are also evolving in response to the explosive growth in data volumes and velocity; While file-based transport remains common, streaming approaches are gaining traction for near real-time processing needs. When file handling is required, modern WMSs must navigate complex storage hierarchies and scale horizontally across distributed storage locations to maintain performance. These trends collectively point toward more sophisticated and flexible systems that can adapt to diverse scientific computing requirements while managing increasingly complex data ecosystems.

\pp{From Extensions to Building Blocks}
As WMSs mature, developers increasingly extend their native capabilities through additional components that address specific needs. This expansion has led to growing system complexity, challenging developers to maintain modular architecture and avoid unwieldy monolithic designs. Rather than each system independently implementing similar functionalities, a promising approach for the workflow community involves identifying and developing shared building blocks, reusable components that provide common services across different WMSs~\cite{hategan2023psi,alsaadi2024exascaleworkflowapplicationsmiddleware,radical}. This community-based approach to the development of modular and interoperable components~\cite{alsaadi2024exascaleworkflowapplicationsmiddleware} could significantly reduce duplication of efforts while improving sustainability and adoption. Such standardized building blocks would address fundamental workflow needs like resource management, data movement, provenance tracking, and fault tolerance, allowing individual systems to focus on their unique strengths and domain-specific optimizations. The emergence of these community-maintained components represents a potential path toward consolidation in a currently fragmented ecosystem of over 300 WMSs, promoting interoperability while preserving the specialized capabilities that particular scientific domains require.

\pp{Workflow Registries in the Scientific Workflow Ecosystem}
In addition to the WMSs themselves, the scientific community has developed various workflow registries that serve as centralized locations to share, discover, and reuse workflow definitions in the workflow ecosystem. These registries complement WMSs by facilitating knowledge exchange and promoting best practices across research domains. The nf-core~\cite{ewels2020nf} repository provides community-maintained curated Nextflow workflows for bioinformatics with continuous integration to ensure reproducibility. Similarly, the SnakeMake workflow catalog~\cite{grayson2023automatic} offers domain-specific collections. WorkflowHub~\cite{gustafsson2025} provides a unified registry for all computational workflows that links to community repositories, making workflows findable, accessible, interoperable, and reusable (FAIR) according to the FAIR principles for workflows~\cite{wilkinson2025}. Unlike single-language workflow registries such as nf-core, the AiiDA plugin registry~\cite{aiida_registry}, and Galaxy Toolshed~\cite{blankenberg2014dissemination} that are associated with specific workflow platforms, WorkflowHub accepts workflows from any scientific domain, in any format and in any workflow language. Repositories such as Dockstore~\cite{yuen2021dockstore} improve reproducibility by combining containers, descriptor languages, and test parameter files to simplify software reuse and dependency management. Dockstore has facilitated large-scale biomedical research collaborations by using cloud technologies to increase the FAIRness of computational resources. WfInstances~\cite{wfcommons} is a key component of the WfCommons project that archives real-world workflow instances collected from workflow executions using various runtime systems. The repository ecosystem represents an important extension of the workflow landscape, bridging technical capabilities with community practices, and helping scientists navigate the complex decision space of workflow selection and reuse while promoting the recognition of workflows as artifacts.

\section{Process to Define the Terminology of Workflow Systems}
\label{sec:process}

The terminology for scientific workflow systems presented in this paper emerged from a systematic, community-driven approach initiated in 2021 through the Workflows Community Initiative (WCI)~\cite{WCI}. This collaborative effort united workflow system developers, domain scientists, and workflow practitioners in diverse scientific disciplines and computing facilities. Through a series of Workflows Community Summit events~\cite{wci-roadmap,da2021workflows, da2021workflows1, da2022workflows, wcs2022, wcs2024}, participants engaged in structured discussions about key aspects of scientific workflow systems, including essential features, challenges in interoperability, data management approaches for execution models and reproducibility requirements. These discussions were documented in technical reports that captured the evolving understanding of WMSs and established the foundations for a unified terminology, drawing inspiration from similar efforts in the in situ processing community~\cite{in-situ-term}.

The development of the terminology progressed through several phases, beginning with an analysis of summit reports and the existing literature on workflow taxonomies~\cite{Yu_taxo05, silva-fgcs17, deelman-fgcs09, liu-grid15, bahsi-2007, kiran2023criteria, Ahmad2021, POOLA2017285, cruz2009} to identify common patterns and classification schemes. A core working group then conceptualized the framework around five distinct axes to comprehensively cover the key aspects of workflows and WMSs, followed by the creation of a draft document defining these axes and their associated terms. This draft included an initial characterization of 23 representative WMSs and was circulated to workflow system developers and key stakeholders for critical feedback. Through multiple iterations of refinement based on community input, the working group adjusted definitions, added missing terms, and ensured that the terminology accurately represented the domain's complexity while remaining both comprehensible and practical. The terminology was validated by applying it to classify the WMSs listed in Table~\ref{tab:workflow_systems}, confirming its applicability while revealing its ability to highlight commonalities and distinctions among diverse systems.

Throughout this process, the working group adhered to the principles of comprehensiveness, accessibility, neutrality, openness, and practicality. The terminology needed to cover the full spectrum of workflow system features without favoring particular implementation approaches, while remaining understandable to both experts and domain scientists. It was designed to be descriptive rather than prescriptive, avoiding implications that certain approaches were inherently superior and flexible enough to accommodate future innovations through the addition of new terms within the established axes. The resulting terminology, as detailed in Section~\ref{sec:axes} and applied in Section~\ref{sec:survey}, represents the collective expertise of a broad community of workflow researchers and practitioners, providing a common language for discussing and comparing WMSs that facilitates both scientific communication and informed decision-making when selecting workflow technologies for specific research needs.

\section{\revised{Related Work}}
\label{sec:related-work}

\revised{
Over the past two decades, the scientific workflow community has proposed several taxonomies to structure the design space of WMSs. Early taxonomies introduced classification schemes based on architectural and infrastructural features, including workflow representation models (e.g., DAGs versus non-DAGs), scheduling strategies (i.e., centralized, decentralized, or hierarchical), fault tolerance mechanisms (e.g., task retries, checkpointing, alternate resource usage), and data movement techniques (e.g., file staging, replication, streaming)~\cite{Yu_taxo05}. These taxonomies highlighted trade-offs between performance, fault resilience, and scalability across grid environments. Later frameworks organized WMS features according to the workflow lifecycle, encompassing composition interfaces (e.g., graphical editors, scripting APIs, domain-specific languages), resource mapping mechanisms (i.e., manual binding versus automated planners), execution engines (i.e., static versus dynamic schedulers), and provenance capture strategies (i.e., retrospective and prospective metadata logging)~\cite{deelman-fgcs09}. These classifications helped emphasize usability and reproducibility as central design goals. More recent comparative analyses have expanded the evaluation criteria to cover support for heterogeneous execution models (including iterative, streaming, and conditionally adaptive workflows), deployment flexibility across HPC, cloud, and hybrid environments, and mechanisms for handling large-scale, data-intensive workloads with performance-aware orchestration and optimized I/O strategies~\cite{silva-fgcs17, kiran2023criteria, Ahmad2021}, or focusing on specific features such as fault tolerance~\cite{POOLA2017285} or provenance~\cite{cruz2009}. Such studies increasingly incorporate practical interoperability, expressiveness, and usability assessments to guide system selection in data-intensive scientific domains.

These taxonomies have provided valuable frameworks for evaluating and selecting WMSs, but they often emphasize either infrastructure-level capabilities or comparisons based on the workflow lifecycle. This paper contributes a complementary approach by proposing a terminology instead of defining yet another hierarchical taxonomy. The objectives are to offer a vocabulary that captures the essential properties of WMSs in a flexible and non-prescriptive manner, support consistent descriptions across heterogeneous systems, and help researchers express requirements and understand systems' capabilities more precisely. Thus, what distinguishes this work is its focus on standardizing language rather than classification alone. By moving away from rigid taxonomies and toward shared terms, we expect to enable clearer communication across domains and stakeholder groups, and support the design, comparison, and integration of next-generation WMSs. Grounded in broad community consensus, the proposed set of terms overlaps with the existing taxonomies, which shows its capacity to capture the main features of classical WMSs. However, it also reflects the evolution of workflow practices, with new terms including dynamic execution behaviors, modular reuse, and serverless orchestration models.
}
\section{Conclusion}
\label{sec:ccl}

In this paper, we have introduced a new terminology for scientific workflow systems. This terminology comprises five axes along which a workflow system can be characterized. Each axis is then refined via multiple associated terms. 
\revised{The development of this terminology is a community-based effort rooted in and supported by the Workflows Community Initiative (WCI). It summarizes the collective thinking of WMS developers and members of the leadership and steering committees of the Workflows Community Initiative and reflects the achieved consensus around an initial set of terms. The main motivation for this work \revised{is} to serve as a starting point for a uniformly understood vocabulary that would help workflow practitioners navigate the vast market of WMSs.} To this end, we used this terminology to characterize a selection of existing WMSs, identify similarities and differences, and highlight some broad trends.  This approach brought in many different perspectives and ensured that diverse perspectives were taken into account. It also provides this terminology with solid foundations and the backing of a significant number of workflow system developers and workflow practitioners. \revised{This will allow us to expose and explain the terminology to the respective user communities of the analyzed frameworks and foster its broader adoption. We also plan to gather and analyze user feedback and monitor the adoption of the terminology to conduct an empirical validation of the benefits of the proposed terminology. A concrete metric of success for the adoption of the terminology will be to be referred to in scientific articles, not by citing this paper but by using the terminology to describe a WMS or a workflow and position contributions using a uniformly understood vocabulary accepted by a broad community.}

\revised{This} terminology should not be considered static. As new systems are developed and new trends emerge from the community, new terms and axes may be introduced\revised{. A new working group of the WCI will be formed, which will include this paper's co-authors to ensure that the terminology evolves and keeps reflecting the state of the field. This group will also be in charge of extending the list of characterized WMSs beyond those that are part of the Workflows Community Initiative.}

\section*{Acknowledgments}
The authors express their deepest appreciation for the insightful review and comments from Khalid Belhajjame of the University Paris-Dauphine (France); Luiz Gadelha of the German Cancer Research Center (DKFZ, Germany); Johan Gustafsson of Australian BioCommons and Sehrish Kanwal of the Centre for Cancer Research at the University of Melbourne (Australia); and Mahnoor Zulfiqar and Stuart Owen of the University of Manchester (United Kingdom).

This research used resources of the Oak Ridge Leadership Computing Facility at the Oak Ridge National Laboratory, supported by the Office of Science of the U.S. Department of Energy under Contract No. DE-AC05-00OR22725. BSC authors acknowledge projects CEX2021-001148-S and PID2023-147979NB-C21 from the  MCIN/AEI and MICIU/AEI /10.13039/501100011033 and by FEDER, UE, and by the Departament de Recerca i Universitats de la Generalitat de Catalunya, research group MPiEDist (2021 SGR 00412).  Ewa Deelman is funded by the U.S. Department of Energy under grant No. DE-SC0024387 and by the U.S. National Science Foundation under grant No. 2138286. This work was performed under the auspices of the US Department of Energy (DOE) by Lawrence Livermore National Laboratory under Contract DE-AC52-07NA27344. This work has been supported by the LDRD at Lawrence Livermore National Laboratory (24-SI-005), LLNL Release number: LLNL-JRNL-2007841. Giovanni Pizzi acknowledges  financial support from the NCCR MARVEL, a National Centre of Competence in Research, funded by the Swiss National Science Foundation (grant number 205602), by the Open Research Data Program of the ETH Board (project ``PREMISE'': Open and Reproducible Materials Science Research) and by the SwissTwins project, funded by the Swiss State Secretariat for Education, Research and Innovation (SERI). Bartosz Balis is funded by the European Union through the Horizon Europe CLOUDSTARS project (101086248).  Douglas Thain acknowledges support from National Science Foundation Grant OCI-2411436. Thain, Chard, Jha, and da Silva acknowledge support from National Science Foundation grant TIP-2346119.

\bibliographystyle{elsarticle-num}

\end{document}